\documentclass[aps,prb,preprint,groupedaddress,nofootinbib]{revtex4-2}
\usepackage{graphicx} 
\usepackage{amsmath}
\usepackage{subfig}
\usepackage{geometry}
\usepackage{setspace}
\usepackage{mdframed}
\usepackage{physics,isomath,amssymb,cleveref} 
\newcommand{\ii}{\mathrm{i}} 
\newcommand{\ee}{\mathrm{e}} 
\newcommand{\vc}{\vb}       
\newcommand{\uv}{\vu}       
\newcommand{\vcbeta}{\vb*{\beta}}       
\newcommand{\uvbeta}{\vu*{\beta}}       
\newcommand{\tens}[1]{\mathsfbfit{{#1}}} 

\usepackage{xcolor}

\date{\today}

\usepackage{etoolbox}

\makeatletter
\patchcmd{\frontmatter@abstract@produce}
  {\vskip200\p@\@plus1fil
   \penalty-200\relax
   \vskip-200\p@\@plus-1fil}
  {}
  {}
  {}
\makeatother

\begin{document}
\singlespacing

\title{Fresnel reflection coefficients in the Fourier domain for a planar surface in uniform motion parallel to its interface.}

\author{St\'ephane Azar}
\affiliation{Department of Physics and London Centre for Nanotechnology, King's College London, Strand, London WC2R 2LS, UK}

\author{Sebastian Golat}
\affiliation{Department of Physics and London Centre for Nanotechnology, King's College London, Strand, London WC2R 2LS, UK}

\author{Francisco~J. Rodr\'iguez-Fortu\~no}
\email{francisco.rodriguez\_fortuno@kcl.ac.uk}
\affiliation{Department of Physics and London Centre for Nanotechnology, King's College London, Strand, London WC2R 2LS, UK}

\begin{abstract}
    The optical reflection coefficient of a dielectric medium moving uniformly in the plane spanned by its surface is rigorously calculated using classical electrodynamics and special relativity, and expressed in the Fourier domain, as a function of the incident frequency and wavevector, valid in both the far- and near-field regimes. It is found that cross-polarisation appears as a consequence of the motion, except when it is directed along the plane of incidence. As an example, using a Drude model for the permittivity of the surface at rest, the dispersion relation of its surface modes is calculated. A tilting of the dispersion relation is observed, leading to movement-induced surface plasmon unidirectionality and non-reciprocity. 
\end{abstract}
\maketitle
\section{Introduction}
 The reflection of light waves on an interface is a well-studied problem, even in the case of moving interfaces. Indeed, moving mirrors have been studied extensively\cite{Sommerfield1954moving_mirror,Pauli1958moving_mirror}. Many different approaches have been used to study this problem, which still has some interesting features to reveal. The first attempt at calculating the reflection on a moving dielectric medium was done in a simplistic 2D model \cite{Yeh1965moving_dielectric}, revealing a Doppler shift based on the direction of motion. This initial attempt was extended by considering motion in different directions and different polarisation of the incident field \cite{Pyati1967moving_dielectric,Shiozawa1967moving_dielectric,Lee1967moving_dielectric,Shiozawa1968moving_dielectric,Huang1994moving_dielectric,Tsai1967moving_media,Idemen2006moving_media,Kunz1980moving_media,Censor1969moving_media,Wang2010moving_dielectric/polarisation_mix}. Then, to go a step further, it was necessary to introduce a model for the electric permittivity. This was mainly done through Drude's model, both lossless and lossy versions \cite{Yeh1966moving_plasma,Yeh1967moving_plasma,Yeh1969moving_plasma,Chawla1969moving_plasma,Mukherjee1973moving_plasma,Chakravarti1971moving_plasma,Gan2020moving_plasma}. From there, many paths were investigated, trying different variations of the original problem and each bringing a new insight. Firstly, different shapes were considered, such as a moving slab \cite{Yeh1966moving_slab,stolyarov1967moving_slab,Yeh1968moving_slab,Yeh1968moving_slab/cylinder/waveguide}, moving cylinder \cite{Lee1967moving_cylinder,Censor1969moving_cylinder,Yeh1968moving_slab/cylinder/waveguide} or moving sphere \cite{Restrick1968moving_sphere,Pogorzelski1973moving_sphere,Radpour2021moving_sphere}. The case of a planar but slightly rough surface was also studied \cite{Chrissoulidis1985rough_surface}. On a different note, instead of changing the shape of the surface, some authors investigated different material properties. Thus, some papers studied the case of moving anisotropic media \cite{Ohkubo1976anisotropic,Besieris1969anisotropic,Mukherjee1976anisotropic}, magnetized media \cite{Mukherjee1974magnetized,Mueller1987magnetized} and inhomogeneous media \cite{Tanaka1972inhomogenous_media}. Other works approached the problem from a more time-dependent perspective, introducing pulses \cite{rattan1973pulse} or even non-uniform motion \cite{Leonhardt1999nonuniformly_moving_media,Tanaka1978accelerating_dielectirc,Tanaka1982accelerating_dielectric}. Studying the problem in a 3D set-up shows that, under motion, the dielectric medium behaves like a magnetoelectric medium \cite{horsley2012polarisation_mix,Wang2010moving_dielectric/polarisation_mix} which are known to produce cross-polarization effects in reflection and transmission \cite{paco2019magnetoelectric}. Moreover, looking at the problem from a quantum mechanical point of view \cite{horsley2012canonical}, it was found that even at $T = 0\,\mathrm{K}$, it is possible to extract energy from the vacuum fluctuations of the electromagnetic field outside of a uniformly moving dielectric. In this paper, we aim to approach the 3D problem of a moving smooth dielectric interface using classical electrodynamics together with special relativity to retrieve the aforementioned results, provide a rigorous and exact expression for the reflection coefficients in the most general linear case, and determine the exact expression for the dispersion relation for surface modes on a moving dielectric. The classical study of this problem uses plane waves and geometrical optics, hence the reflection coefficient is calculated in terms of angles of incidence, and not wavevectors. We wish to find the reflection coefficient in the Fourier domain, in terms of the incident frequency and wavevector, explicitly allowing values outside the light cone. This approach greatly simplifies the expressions and the simplicity in the derivation, and more importantly, our expressions for the reflection coefficients are designed to be fully compatible with Fourier-based methods such as angular spectrum approaches and Green function calculations, enabling the study of not only propagating waves but also near-field sources in close proximity to the surface. For instance, the introduction of evanescent waves in the picture is capital to study the dispersion relation of surface modes, which was lacking in previous papers. 
\section{Reflection coefficients}
\subsection{General Derivation}
\begin{figure}[ht]
    \centering
    \includegraphics[width=8.5cm]{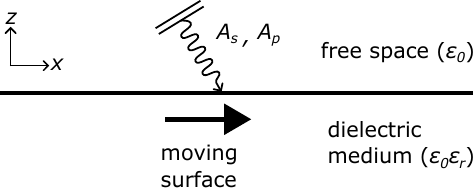}
    \caption{Diagram of an incident field on the moving dielectric medium surface.}
    \label{moving surface}
\end{figure}
We start by deriving the reflection coefficients of a moving dielectric medium. 
Let's consider a dielectric medium spanning the half-space $z < \text{const}$, with free space above ($z > \text{const}$). This dielectric is allowed to move in any direction parallel to the interface with a velocity $\mathbf{v}$. We orient our axis so that the surface motion is along the $x$-axis, giving $\mathbf{v} = v\mathbf{\hat{x}} $ (see Fig. \ref{moving surface}). 
Any incident field can be decomposed in the temporal and spatial Fourier domains into a sum of homogeneous waves. An incident homogeneous wave can be defined as having a fixed angular frequency $\omega$ and a wavevector $\mathbf{k}^{\pm} = k_x \mathbf{\hat{x}}+k_y \mathbf{\hat{y}}\pm k_z \mathbf{\hat{z}}$. We ensure throughout our derivation that both the frequency and the wavevector components may, or may not, have a complex value. The variables of our problem on which the reflection coefficient will depend are the transverse wavevectors $k_x$, $k_y$ and the frequency $\omega$, these three variables are conserved due to the translational invariance in the XY plane and the time-translation invariance of the problem. The $z$-component of the wavevector $k_z$, in contrast, is a dependent variable, determined by the dispersion relation in vacuum $k_x^2+k_y^2+k_z^2=k_0^2=\omega^2/c^2$. We will use the notation $k_z(k_x,k_y,\omega)=+\sqrt{k_0^2-k_x^2-k_y^2}$ to always be the positive square root solution (in the half-open range $0\leq\arg{k_z}<\pi$), and use the $\pm$ in $\mathbf{k}^{\pm}$ to determine if the plane/evanescent wave propagates/decays along the positive or negative $z$ direction. The electric and magnetic fields of this single wave can always be decomposed into two polarisation components, corresponding to the two complex degrees of freedom of electromagnetic waves. The most common choice of polarisation basis is known as $s$ and $p$ polarisation, in which the electric field is either perpendicular or parallel to the plane of incidence, which is the plane containing the incident wavevector and perpendicular to the plane of the dielectric interface:
\begin{equation}
\begin{aligned}
    \label{eq:incidentplanewave}
    \mathbf{E}(\mathbf{r},t)&= \left( A_s \mathbf{\hat{e}}_s+A_p \mathbf{\hat{e}}_p^-\right) e^{i\mathbf{k}^{-}\cdot\mathbf{r}-i \omega t }\\
    \mathbf{B}(\mathbf{r},t)&=\frac{1}{c}\left(A_p \mathbf{\hat{e}}_s-A_s \mathbf{\hat{e}}_p^-\right) e^{i\mathbf{k}^{-}\cdot\mathbf{r}-i \omega t }
\end{aligned}
\end{equation}
\noindent 
with the polarisation unit vectors defined, in free space, as:
%
%
\begin{equation}
\label{eq:basisvectors}
\mathbf{\hat{e}}_s(\mathbf{k},\omega) = \frac{\mathbf{\hat{z}}\times\mathbf{k}^{\pm}}{\sqrt{\left( \mathbf{\hat{z}}\times\mathbf{k}^{\pm} \right)\cdot \left( \mathbf{\hat{z}}\times\mathbf{k}^{\pm} \right)}} =\frac{1}{k_t} \begin{pmatrix}
-k_y\\
 k_x\\
 0\\
\end{pmatrix}  
\hspace{2cm}
\mathbf{\hat{e}}_p^\pm(\mathbf{k},\omega) =\mathbf{\hat{e}}_s\times \frac{\mathbf{k}^{\pm}}{k_0}=\frac{1}{k_t k_0} \begin{pmatrix}
\pm k_xk_z\\
 \pm k_yk_z\\
 - k_t^2\\
\end{pmatrix}
\end{equation}
\noindent
with $k_t = \sqrt{k_x^2 + k_y^2}$ being the transverse wavevector. For a plane wave with a purely real $\mathbf{k}$-vector, these two unit vectors constitute two transverse linear polarisations. They correspond to the polar $\mathbf{\hat{e}}_p = \mathbf{\hat{e}}_\theta$ and azimuthal $\mathbf{\hat{e}}_s = \mathbf{\hat{e}}_\phi$ spherical coordinate unit vectors, defined in $k$-space with the $\mathbf{k}$-vector as the radial direction. Crucially, however, the mathematical description above remains valid for any $\mathbf{k}$-vector, including evanescent waves, where the polarisations become elliptical, non-transverse, and three-dimensional. In that case, the unit vectors are complex-valued, but fulfil the transversality condition $\mathbf{\hat{e}}_{p}^\pm \cdot \mathbf{k}^\pm = \mathbf{\hat{e}}_{s} \cdot \mathbf{k}^\pm = 0$ required by Maxwell's equations in free space, and the orthonormality of the basis $\mathbf{\hat{e}}_{p}^\pm \cdot \mathbf{\hat{e}}_{s} = 0$, and $ \mathbf{\hat{e}}_{p}^\pm \cdot \mathbf{\hat{e}}_{p}^\pm = \mathbf{\hat{e}}_{s} \cdot \mathbf{\hat{e}}_{s} = 1$, where no complex conjugation is needed. The $\pm$ sign in $\mathbf{\hat{e}}_p^\pm$ accounts for the two possible signs of $k_z$. For the incident wave coming towards the surface, in the negative $z$ direction, we must take the $(-)$ sign.
%
%
%
%
%
Now this wave will reflect off the moving surface. To rigorously calculate the reflection coefficient, we first need to write the incident plane wave from the point of view of the surface (i.e. in a frame moving with the surface, in which the reflection coefficients are known). To do this, we turn to the relativistic notation of electromagnetism. We construct the electromagnetic field tensor as \cite{Griffiths2012}:

\begin{equation}
\label{eq:EMtensor}
    F^{\mu\nu} = \begin{pmatrix}
0 & \frac{E_x}{c}&\frac{E_y}{c}&\frac{E_z}{c}\\
 -\frac{E_x}{c}& 0&B_z &-B_y\\
 -\frac{E_y}{c}&-B_z &0 &B_x\\
 -\frac{E_z}{c}&B_y &-B_x &0\\
\end{pmatrix}
\end{equation}
Furthermore, the 4-wavevector is given as $k^{\mu} = \begin{pmatrix}
\frac{\omega}{c}& k_x& k_y& k_z
\end{pmatrix}^T$ , where $\left(\frac{\omega}{c}\right)^2 = k_0^2 = k_x^2 + k_y^2 +k_z^2$. Concerning the surface, since it is moving along the $x$-axis with a velocity  $\mathbf{v} = v\mathbf{\hat{x}}$ in the lab frame, we can get to the frame where the surface  is at rest (denoted by a prime) using a Lorentz boost of the form \cite{Griffiths2012}:
\begin{equation}
\label{eq:Lorentz_boost_matrix}
    \Lambda_\alpha^{\mu} = \begin{pmatrix}
 \gamma&-\gamma\beta&0&0\\
 -\gamma\beta&\gamma&0&0\\
 0&0&1&0\\
 0&0&0&1\\
\end{pmatrix}
\end{equation}
where $\beta = v/c$ is the relative speed factor, and $\gamma = (1 - \beta^2)^{-1/2}$ is the Lorentz factor. Using this boost, the transformed 4-wavevector and the transformed electromagnetic field 4-tensor are given by:
\begin{equation}
\begin{aligned}
    k'^{\rho \pm} &= \Lambda_\mu^{\rho}k^{\mu \pm}, \\
    F'^{\mu\nu} &= \Lambda_\alpha^{\mu}\Lambda_\beta^{\nu}F^{\alpha\beta}.
    \label{eq:Lorentz_boost}
\end{aligned}
\end{equation}
%
%
%
%
The first equation gives us the frequency and wavevector of the incident field as seen from the reference frame of the surface:
\begin{equation}
\begin{aligned}
    \label{eq:transformation_wavevector}
    \omega' = \gamma \omega - \gamma \beta c k_x, \qquad
    k_x' = \gamma k_x -\gamma \beta \omega/c, \qquad
    k_y' = k_y, \qquad k_z' = k_z
\end{aligned}
\end{equation}
while the second equation allows us to obtain the incident electric field vector $\mathbf{E}_{\text{inc}}'$, as seen from the reference frame of the surface, which we can expand into $s$ and $p$ components in this frame, with the unit vectors now depending on $(\omega',\mathbf{k}')$, i.e., on $k'^\mu$:
\begin{equation}
\begin{aligned}
\label{eq:incidentplanewaveboostedframe}
\mathbf{E}_{\text{inc}}'(\mathbf{r}',t')&=\left( A_{\text{inc},s}' \mathbf{\hat{e}}_s(\omega',k_x',k_y')+A_{\text{inc},p}' \mathbf{\hat{e}}_p^-(\omega',k_x',k_y')\right) e^{i\mathbf{k}'^{-}\cdot\mathbf{r}'-i \omega' t' }
\end{aligned}
\end{equation}
In accordance to the above equation, the $s$ and $p$ component amplitudes in the boosted frame can be retrieved by exploiting the orthonormality of the polarisation basis, by doing a dot product with the unit vectors evaluated in the same frame:
\begin{equation}
\label{eq:Aspdotproduct}
\begin{aligned}
    A_{\text{inc},s}' &= \mathbf{E}_{\text{inc}}'(\mathbf{r}'=0,t'=0) \cdot \mathbf{\hat{e}}_s(\omega',k_x',k_y') \\
    A_{\text{inc},p}' &= \mathbf{E}_{\text{inc}}'(\mathbf{r}'=0,t'=0) \cdot \mathbf{\hat{e}}_p^-(\omega',k_x',k_y') 
\end{aligned}
\end{equation}
%
Doing the exact algebra from eq.~(\ref{eq:Aspdotproduct}) after combining eqs.~(\ref{eq:incidentplanewave}--\ref{eq:Lorentz_boost}), we can express the incident polarisation in the boosted frame in terms of the polarisation in the lab frame as follows:
\begin{equation}\label{eq:boost_of_Jones}
    \begin{pmatrix}
        A_{\text{inc},s}'\\
        A_{\text{inc},p}'
    \end{pmatrix} = \frac{1}{k_tk_t'}\begin{pmatrix}
         k_t'^2 + \gamma\beta \frac{k_x'k_z^2 }{k_0} &  \gamma\beta \frac{k'_0k_yk_z}{k_0}\\
           -\gamma\beta \frac{k'_0k_yk_z}{k_0} &  k_t'^2 + \gamma\beta \frac{k_x'k_z^2}{k_0} \\
    \end{pmatrix}  \begin{pmatrix}
        A_s\\
        A_p
    \end{pmatrix}    
\end{equation}
Note that this ``boosting" matrix takes into account the fact that our incident wave is propagating down through the negative superscript in Eqs.~(\ref{eq:incidentplanewaveboostedframe}) and (\ref{eq:Aspdotproduct}) which accounts for a negative sign in front of every appearance of $k_z$. 
In the rest frame of the surface, the reflection of the plane wave is well known, involving the usual Fresnel reflection coefficients. In the most general linear case, the surface at rest has a reflection matrix of the form: 
\begin{equation}
\begin{aligned}
    \mathbf{R}_0(\omega',k_x',k_y') &= \begin{pmatrix}
    r_{ss0}(\omega',k_x',k_y') & r_{sp0}(\omega',k_x',k_y') \\
    r_{ps0}(\omega',k_x',k_y') & r_{pp0}(\omega',k_x',k_y')
    \end{pmatrix}
    \end{aligned}
\end{equation}
Such that we can calculate the reflection in the frame of the surface:
\begin{equation}
    \begin{pmatrix}
        A_{\text{ref},s}'\\
        A_{\text{ref},p}'
    \end{pmatrix} =   \mathbf{R}_0(\omega',k_x',k_y')\begin{pmatrix}
    A_{\text{inc},s}'\\
        A_{\text{inc},p}'
    \end{pmatrix}
\end{equation}
The next step is to boost the reflected field back into the lab frame. For that, we follow the same steps that led to eq.~(\ref{eq:boost_of_Jones}) with a few considerations. Since we are boosting from the frame of the surface to the lab frame we need to interchange all primed and unprimed arguments, which comes down to $k_x \leftrightarrow k_x'$,  $k_0 \leftrightarrow k_0'$ and  $k_t \leftrightarrow k_t'$. Furthermore, this boost now has a velocity of $\mathbf{v}' = -v\hat{\mathbf{x}}$ so we also need to change the sign $\beta \rightarrow -\beta$. In addition, the reflected field is now propagating upward which means we must take the $(+)$ superscript sign when rewriting eqs.~(\ref{eq:incidentplanewaveboostedframe}) to (\ref{eq:Aspdotproduct}) for the reflected plane waves. This change is equivalent to doing $-k_z \rightarrow k_z$. With this we get the following expressions to relate the reflected field polarisation in the boosted frame to the one in the lab frame.
\begin{equation}
    \label{eq:inverse_boost_matrix}
    \begin{pmatrix}
        A_{\text{ref},s}\\
        A_{\text{ref},p}
    \end{pmatrix} = \frac{1}{k_tk_t'}\begin{pmatrix}
         k_t^2 - \gamma\beta \frac{k_xk_z^2}{k_0'} &  \gamma\beta \frac{k_0k_yk_z}{k_0'}\\
           -\gamma\beta \frac{k_0k_yk_z}{k_0'} &  k_t^2 - \gamma\beta \frac{k_xk_z^2}{k_0'}\\
    \end{pmatrix}  \begin{pmatrix}
    A_{\text{ref},s}'\\
        A_{\text{ref},p}'
    \end{pmatrix}
\end{equation}
Putting everything together [eqs.~(\ref{eq:boost_of_Jones}--\ref{eq:inverse_boost_matrix})], we are able to express the reflected field as a function of the incident field, all expressed in the lab frame in which the surface is moving:
\begin{equation}
    \begin{pmatrix}
        A_{\text{ref},s}\\
        A_{\text{ref},p}
    \end{pmatrix} = \mathbf{R}(\omega,k_x,k_y)  \begin{pmatrix}
        A_s\\
        A_p
    \end{pmatrix}
\end{equation}
\noindent where the reflection coefficient matrix $\mathbf{R}(\omega,k_x,k_y)$ is given by:
\begin{mdframed}[linecolor=black!30!green!40,backgroundcolor=green!2, linewidth=1pt]%
\vspace{-1em}
\begin{equation}
    \label{eq:final_general_reflection_matrix}
    \mathbf{R}(\omega,k_x,k_y) = \frac{1}{k_t^2k_t'^2}\begin{pmatrix}
         k_t^2 - \gamma\beta \frac{k_xk_z^2}{k_0'} &  \gamma\beta \frac{k_0k_yk_z}{k_0'}\\
           -\gamma\beta \frac{k_0k_yk_z}{k_0'} &  k_t^2 - \gamma\beta \frac{k_xk_z^2}{k_0'}\\
    \end{pmatrix} \mathbf{R}_0(\omega',k_x',k_y') \begin{pmatrix}
         k_t'^2 + \gamma\beta \frac{k_x'k_z^2}{k_0} &  \gamma\beta \frac{k_0'k_yk_z}{k_0}\\
           -\gamma\beta \frac{k_0'k_yk_z}{k_0} &  k_t'^2 + \gamma\beta \frac{k_x'k_z^2}{k_0}\\
    \end{pmatrix}
\end{equation}
\end{mdframed}

This, together with the relations for $(\omega',k_x',k_y')$ in terms of $(\omega,k_x,k_y)$ given in eq.~(\ref{eq:transformation_wavevector}), and remembering $\beta = v/c$, $\gamma = (1 - \beta^2)^{-1/2}$, $k_0=\omega/c$ and $k_0'=\omega'/c$, form a rigorous final expression for the reflection matrix of a smooth interface between free space and some other arbitrary material in uniform motion parallel to its surface, with $\mathbf{R}_0(\omega,k_x,k_y)$ being the reflection matrix of that same interface when at rest. This is the main result of this work, it is an exact solution within the context of classical electrodynamics and special relativity, and has the following advantages: the reflection matrix of the surface at rest is kept as a variable $\mathbf{R}_0$, therefore encompassing any linear surface including, for instance, anisotropic or magneto-optical materials, as well as slabs. Also, the reflection matrix $\mathbf{R}$ depends on the incident wavevector components and frequency, with no mention of angles, and is valid for both propagating and evanescent waves, and for complex values in $\omega$, $k_x$ and/or $k_y$, being well-suited to the standard mathematical tools used in nanophotonics problems, and directly applicable to the Fourier transform of any electromagnetic field.
\subsection{Simple Dielectric case with no initial cross-polarization}
Up to this point, we made no specific assumption on the reflection coefficient of the surface when it is at rest. We now consider it to be an isotropic dielectric medium with relative permittivity $\varepsilon_r$, which implies no cross-polarization terms $r_{sp0}$ and $r_{ps0}$ initially. Using eq.~(\ref{eq:final_general_reflection_matrix}) derived above, it is now possible to calculate the exact form of the reflection coefficient in the lab frame.
\begin{equation}
\label{eq:reflection_simple_dielectric}
    \begin{aligned}
    \mathbf{R}(\omega,k_x,k_y) &= \begin{pmatrix}
        r_{ss} & r_{sp} \\
        r_{ps} & r_{pp}
        \end{pmatrix} \\
    r_{ss}(\omega,k_x,k_y) &= \frac{1}{k_t^2k_t'^2} \left[ (k_t^2k_t'^2+ (\gamma\beta k_yk_z)^2)r_{s0} - (\gamma\beta k_yk_z)^2r_{p0} \right] \\
    r_{sp}(\omega,k_x,k_y) &= \frac{1}{k_t^2k_t'^2} \left( \gamma\beta k_yk_z\left[ \frac{k_0'}{k_0}(k_t^2 - \gamma\beta \frac{k_xk_z^2}{k_0'})r_{s0}+\frac{k_0}{k_0'}(k_t'^2 + \gamma\beta \frac{k_x'k_z^2}{k_0})r_{p0}\right] \right) \\
    r_{ps}(\omega,k_x,k_y) &= \frac{1}{k_t^2k_t'^2} \left(  -\gamma\beta k_yk_z \left[\frac{k_0}{k_0'}(k_t'^2 + \gamma\beta \frac{k_x'k_z^2}{k_0})r_{s0}+\frac{k_0'}{k_0}(k_t^2 - \gamma\beta \frac{k_xk_z^2}{k_0'})r_{p0}\right] \right) \\
    r_{pp}(\omega,k_x,k_y) &= \frac{1}{k_t^2k_t'^2} \left[ (k_t^2k_t'^2+ (\gamma\beta k_yk_z)^2)r_{p0} - (\gamma\beta k_yk_z)^2r_{s0} \right]
    \end{aligned}
\end{equation}
This expression clearly shows us that, even though the surface at rest doesn't exhibit any cross-polarization, once it moves in space it introduces cross-polarization, as observed in \cite{horsley2012polarisation_mix}. However, as mentioned in \cite{horsley2012polarisation_mix} the cross terms are linear in $\gamma\beta$ so they become noticeable only at relatively high speed.
\subsection{2D case: restricting to the plane of incidence}
Much simpler expressions can be obtained in a two-dimensional problem in which the surface moves along the plane of incidence of the incident fields. In our case, since we already defined our surface motion to be along the $x$ direction, we simply restrict the plane of incidence to the XZ plane. Therefore, the requirement is simply that $k_y = 0$. Indeed, any field that is invariant in the $y$-direction can be expressed in the Fourier domain with all its components fulfilling $k_y = 0$. With that, the reflection coefficient becomes extremely simple, as it reduces to:
\begin{equation}
    \mathbf{R}(\omega,k_x) = \begin{pmatrix}
    r_{s0}(\omega',k_x') & 0 \\
    0 & r_{p0}(\omega',k_x')
\end{pmatrix}
\end{equation}
where the only difference with respect to the surface at rest lies in the change of frequency and wavevector incurred by the Lorentz boost. As remarked in \cite{horsley2012polarisation_mix}, when the motion of the surface is along the plane of incidence, the mixing of polarization disappears. Nonetheless, it remains interesting to study this case since we can now focus on how the reflection coefficients change individually.
Using the known equations for the reflection coefficients of a simple dielectric medium \cite{Novotny2011} and substituting in the primed variables by their expression in eq.~(\ref{eq:transformation_wavevector}), we arrive at: 
\begin{equation}
    r_s(\omega,k_x) = \frac{k_z - \sqrt{(\varepsilon_r-1)[\gamma k_x - \gamma \beta k_0]^2  +\varepsilon_r k_z^2}}{k_z + \sqrt{(\varepsilon_r-1)[\gamma k_x - \gamma \beta k_0]^2 +\varepsilon_r k_z^2}}
\end{equation}

\begin{equation}
\label{eq:rp_2d_isotropic_problem}
    r_p(\omega,k_x) = \frac{\varepsilon_r k_z - \sqrt{(\varepsilon_r-1)[\gamma k_x - \gamma \beta k_0]^2+\varepsilon_r k_z^2}}{\varepsilon_r k_z + \sqrt{(\varepsilon_r-1)[\gamma k_x - \gamma \beta k_0]^2  +\varepsilon_r k_z^2}}
\end{equation}

This equation for $r_s$ is equivalent to the formula found by Yeh in \cite{Yeh1965moving_dielectric}, where only $s$-polarization was considered. One key feature to remark is that, at rest, the reflection coefficients are circularly symmetric with respect to $(k_x,k_y)$ since they only depend on $k_t^2$, but here it is no longer the case since $r_{s,p}(\omega,k_x) \neq r_{s,p}(\omega,-k_x) $. This means that the movement of the surface breaks the mirror symmetry of the problem, and there will an asymmetry in the reflected fields depending on the speed of the surface. However, if we consider $\beta$ as a variable, we notice that $r_{s,p}(\beta,k_x) = r_{s,p}(-\beta,-k_x) $, consolidating the idea that the asymmetry in reflection is relative to the speed of the surface. 
\vspace{-1.5ex}
\section{Dispersion relation of surface plasmons on a moving metal} 
\vspace{-1ex}
Using the expression we found for the reflection coefficients, and knowing these are valid for all incident wavevectors including evanescent ones, we are able to determine the dispersion relation of the surface plasmons on a moving metallic interface. We shall focus on $p$-polarization in this section as it is known that $s$-polarized light doesn't produce plasmons.
In order to find an equation for the dispersion relation followed by self-sustaining modes on the surface, we may set the denominator of $r_p(\omega,k_x)$ to $0$ representing the idea that a self-sustaining mode exists in this geometry even in the absence of incident light. This can be done directly from the above equation, or we can start from the fact that, in this case, $r_p(\omega,k_x) = r_{p0}(\omega'(\omega, k_x),k_x'(\omega, k_x))$.  Then it is easier to solve for $k_{\text{spp}}'$ since $r_{p0}$ is the usual reflection coefficient for which we already know the dispersion relation (zeros in the denominator) to be given by \cite{Maier2007plasmonic}: 
\begin{equation}
    k_\text{spp}'^\pm = \pm \sqrt{\frac{\varepsilon_r(\omega')}{\varepsilon_r(\omega') + 1}} \hspace{0.1cm} k_0' 
\end{equation}
From which we only need to apply the relations given by the Lorentz boost (eq. \ref{eq:transformation_wavevector}) to arrive at an expression for $k_\text{spp}$ in the lab frame.
\begin{equation}
    k_\text{spp}^\pm = \frac{\pm \sqrt{\varepsilon_r(\omega')} + \beta \sqrt{\varepsilon_r(\omega') + 1}}{\sqrt{\varepsilon_r(\omega') + 1} \pm \beta \sqrt{\varepsilon_r(\omega')}} \hspace{0.1cm} k_0
\end{equation}
\begin{figure}[h]
    \centering
    \includegraphics[width=17cm]{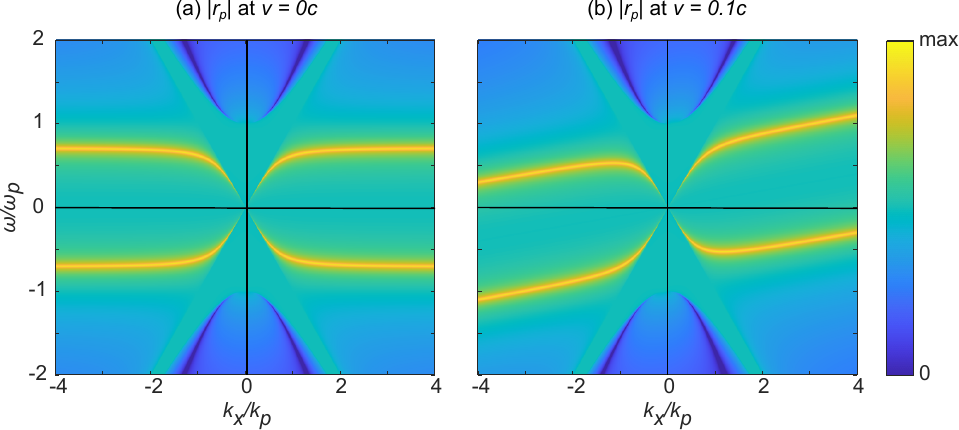}
    \caption{Plot of the absolute value of the reflection coefficient $r_p$, on a logarithmic scale, as a function of $\omega/\omega_p$ and $k_x/k_p$, where $\omega_p$ is the plasma frequency and $k_p = \omega_p/c$. The dispersion relation of the surface corresponds to the maximum of $r_p$ shown by the yellow lines. With (a) the surface being at rest and (b) moving uniformly.}
    \label{dispersion relation}
\end{figure}
However, we notice that the dielectric permittivity $\varepsilon_r$ depends on $\omega'$ which itself depends on $k_x$ [in accordance to eq.~(\ref{eq:transformation_wavevector})] meaning that this is an implicit equation for $k_\text{spp}$. To go further we need to chose a model for the electric permittivity of our surface. Here we will consider a lossless Drude model, typical of plasmonic metals, such that  $\varepsilon_r(\omega') = 1 - \omega_p^2/\omega'^2$. 
With this we can now plot the dispersion relation at any given speed. There are many possibilities to plot the dispersion relation, the first being to search for the minimum of $|k_x - k_{\text{spp}}^\pm|$, which simply solves the above equation numerically. However, it is much simpler to directly plot $r_p$ or its denominator from eq.~(\ref{eq:rp_2d_isotropic_problem}) and to look for the maximum of the former or the minimum of the latter.
Fig.~\ref{dispersion relation} shows the reflection coefficient as a function of real $\omega$ and $k_x$. The yellow lines are maxima of $r_p$ which correspond to the dispersion relation. In this case, as the material was lossless, the surface mode dispersion curves lie on the real $\omega$ and $k_x$ plane. Lossy materials will move the dispersion relation curves into complex values of $\omega$ or $k_x$, but as remarked previously, our expressions remain valid in that case. The most striking feature seen in Fig.~\ref{dispersion relation} is that it is a tilted version of the dispersion relation of a dielectric medium at rest. This tilting introduces many features concerning the plasmons. One of which is that for a high enough frequency only a single direction of surface plasmon is allowed, implying a movement-based surface plasmon unidirectionality, similar to the electric-current-based directionality suggested in \cite{paco2018current_induced_plasmons}. 
\section{Conclusion}
In this paper, we provide a simple clean derivation for the reflection coefficients, and therefore the reflected fields, from a uniformly moving dielectric medium. Our expressions are given in the Fourier domain, allowing a simple use of the angular spectrum representation of fields to study the whole spectrum of wavevectors, including both propagating and evanescent waves. Our expressions are designed and perfectly suited to be used in well-known nanophotonic mathematical techniques, such as Green's function approaches, with direct application in problems such as near field excitation of plasmons, density of states calculations, or even Casimir force calculations and thermal emission calculations based on integrations along a Fourier domain.  Our results are very general, and match the results of previous authors under their respective assumptions. We confirmed that cross-polarisation is introduced by the motion of the surface, except when the motion is directed along the plane of incidence, in agreement with Ref.~\cite{horsley2012polarisation_mix}. Furthermore, our derivation was specifically designed to be fully valid in the evanescent region of the spectrum, allowing us to study the dispersion relation of this moving surface, which tilts along the direction of motion, confirming a movement-induced surface plasmon unidirectionality and non-reciprocity. Furthermore, this tilting of the dispersion relation presents new interesting surface plasmon polariton regimes to excite on the surface, suitable for further study. Indeed, the presented method used to calculate the reflection coefficient can be easily reapplied in different set-ups such as a moving slab \cite{Yeh1966moving_slab,stolyarov1967moving_slab,Yeh1968moving_slab,Yeh1968moving_slab/cylinder/waveguide}. This can be introduced into our expression by substituting the known reflection matrix $\mathbf{R}_0$ for a stationary slab. Transmission coefficients can be easily obtained with the same methods and very few changes. By being expressed in the Fourier domain, a slab calculation would be compatible with the angular spectrum approach, allowing direct application to transfer-matrix methods that can study the interaction of far and near fields with moving slabs. Finally, the reflection coefficient of a moving surface can form part of more complex situations, such as a metal-insulator-metal waveguide with both walls undergoing independent motion.
\section*{Acknowledgments}
S. A. is supported by EPSRC through an NMESFS-2024 scholarship. S. G. and F.J. R.-F. were supported by European Innovation Council (EIC) Pathfinder 101046961 CHIRALFORCE.
\section*{Authors Contribution}
S. A. performed the analytical derivations under the supervision of F.J. R.-F. The general calculation in the appendix was done by S. G. The writing of the manuscript and creation of figures was mainly done by S. A. with supervision, contributions and feedback from all authors.
\appendix
\section{Boost of Jones vector in an arbitrary direction}
In the main text, we considered boosts along the $x$-direction. In this appendix we consider boosts along any arbitrary direction, requiring an understanding of how the Jones vector of polarisation transforms for any plane or evanescent wave under arbitrary Lorentz boosts. This appendix formalises that transformation.

Consider the action of a Lorentz boost on the components of 4-vector $k^\mu$, which can be represented as a matrix:
\begin{equation}
    {k'^\mu}=\pmqty{k'_0\\\vc{k}'}=
    \pmqty{\gamma&-\gamma\vcbeta^\intercal\\-\gamma\vcbeta&\vc{\Gamma}_{\parallel}(\vcbeta)}
    \pmqty{k_0\\\vc{k}}={\Lambda^\mu}_\nu k^\nu
\end{equation}
where $\vcbeta=\beta\uvbeta$ {is the relative velocity of the boost, with $\beta=v/c$ and} $\uvbeta$ being the direction of the boost, and we have defined a parallel stretching operator
\begin{equation}\label{eq:stretch_parallel}
    \vc{\Gamma}_\parallel(\vcbeta)=
            (\hat{I}-\uvbeta\uvbeta^\intercal)+\gamma\uvbeta\uvbeta^\intercal=\tens{P}_\perp(\uvbeta)+\gamma\tens{P}_\parallel(\uvbeta)\,,
\end{equation}
where $\hat{I}=\operatorname{diag}(1,1,1)$, and $\tens{P}_\perp$ and $\tens{P}_\parallel$ are transverse and longitudinal projection operators. This transformation works for the frequency and wavevector, however, the electromagnetic field is described by an antisymmetric rank-2 tensor, which transforms in the following way:
\begin{equation}
    {F'^\mu}_\nu={\Lambda^\mu}_\rho {F^\rho}_\sigma {\Lambda^\sigma}_\nu={(\Lambda F \Lambda^{-1})^\mu}_\nu=(\operatorname{Ad}_{\Lambda}F)^{\mu}{}_{\nu}
\end{equation}
where $\operatorname{Ad}_{\Lambda}(X)=\Lambda X \Lambda^{-1}$ is called the adjoint representation of the Lorentz group. Since $F^{\mu\nu}$ is an antisymmetric 4-by-4 matrix with six independent components, it is isomorphic to a 6-vector $\pmqty{\vc{E}/c,\vc{B}}^\intercal$ and the adjoint representation $\operatorname{Ad}_{\Lambda}$ can then be understood as a 6-by-6 matrix acting as follows:
\begin{equation}
\vec{F}'=\pmqty{\vc{E}'/c\\\vc{B}'}=\pmqty{\vc{\Gamma}_\perp(\vcbeta)&\gamma\vcbeta\cp\\-\gamma\vcbeta\cp&\vc{\Gamma}_\perp(\vcbeta)}\pmqty{\vc{E}/c\\\vc{B}}=\operatorname{Ad}_{\Lambda}\vec{F}\,,
\end{equation}
where we use arrows for 6-vectors, and we introduced the transverse stretching operator defined as:
\begin{equation}\label{eq:stretch_perp}
    \vc{\Gamma}_\perp(\vcbeta)=
            \gamma(\hat{I}-\uvbeta\uvbeta^\intercal)+\uvbeta\uvbeta^\intercal=
        \gamma\tens{P}_\perp(\uvbeta)+\tens{P}_\parallel(\uvbeta)\,.
\end{equation}
Notice that these 6-vectors can be understood as 2-vectors (separate electric and magnetic field) with components that are themselves 3-vectors (their spatial components), and these can be treated separately.\footnote{This is because they live in a vector space $\mathbb{R}^6=\mathbb{R}^2\otimes\mathbb{R}^3$ (or $\mathbb{C}^6=\mathbb{C}^2\otimes\mathbb{C}^3$ for analytic signals or phasors).} This also means that one can write the matrix $\operatorname{Ad}_{\Lambda}$ in terms of the tensor product of Pauli matrices acting on the 2-vectors (separating $\vc{E}$ and $\vc{H}$ contributions) and 3-by-3 matrices acting on 3-vector components (so $\uv{x}$, $\uv{y}$ ,$\uv{z}$ components):
\begin{equation}\label{eq:boost}
    \operatorname{Ad}_{\Lambda}(\vcbeta)=\vc{\Gamma}_\perp(\vcbeta)\pmqty{\pmat{0}}+\ii\gamma\vcbeta\cp \pmqty{0&-\ii\\\ii&0}=\vc{\Gamma}_\perp(\vcbeta)\tens{\sigma}_0+\ii\gamma\vcbeta\cp\tens{\sigma}_2\,,
\end{equation}
where we keep the tensor product implicit.\footnote{The symbol $\vcbeta \times$ denotes the antisymmetric matrix representing the cross product operator. That is, $\vcbeta \times \vc{a} \equiv [\vcbeta \times]\, \vc{a}$ for any vector $\vc{a}$, where $[\vcbeta \times]$ is the $3 \times 3$ matrix:
\[
[\vcbeta \times] = \pmqty{
0 & -\beta_z & \beta_y \\
\beta_z & 0 & -\beta_x \\
-\beta_y & \beta_x & 0
}.
\]
}
The plane wave fields \cref{eq:incidentplanewave} can be written in this representation as:
\begin{equation}
    c\vec{F}(t,\vc{r})=\pmqty{\vc{E}(t,\vc{r})\\c\vc{B}(t,\vc{r})}=\underbrace{\pmqty{A_s\uv{e}_s+A_p\uv{e}_p\\A_p\uv{e}_s-A_s\uv{e}_p}}_{c\vec{F}_0(\omega,\vc{k})}\ee^{\ii\vc{k}\vdot\vc{r}-\ii\omega t}
    =
    \underbrace{\pmqty{\uv{e}_s&\uv{e}_p\\-\uv{e}_p&\uv{e}_s}}_{\tens{M}(\omega,\vc{k})}\underbrace{\pmqty{A_s\\A_p}}_{\vb*{\mathcal{J}}\vphantom{\tens{M}(\omega,\vc{k})}}\ee^{\ii\vc{k}\vdot\vc{r}-\ii\omega t}\,,
\end{equation}
where in the last step, we wrote it in terms of the familiar Jones vector of polarisation $\vb*{\mathcal{J}}$ (using bold calligraphic to emphasise it is a 2-vector). The linear map $\tens{M}:\mathbb{C}^2\to \mathbb{C}^6$ such that $ \vb*{\mathcal{J}}\mapsto c\vec{F}_0(\omega,\vc{k})=\tens{M}(\omega,\vc{k})\vb*{\mathcal{J}}$ might seem a bit strange at first, but it is well defined. For instance, it can be represented by a rectangular matrix:
\begin{equation}
    \tens{M}\vb*{\mathcal{J}}=\underbrace{\pmqty{\uv{e}_s&\uv{e}_p\\-\uv{e}_p&\uv{e}_s}}_{6\times2}\underbrace{\pmqty{A_s\\A_p}}_{2\times1}=\pmqty{
    \uv{x}\vdot\uv{e}_s&\uv{x}\vdot\uv{e}_p\\
    \uv{y}\vdot\uv{e}_s&\uv{y}\vdot\uv{e}_p\\
    \uv{z}\vdot\uv{e}_s&\uv{z}\vdot\uv{e}_p\\
    -\uv{x}\vdot\uv{e}_p&\uv{x}\vdot\uv{e}_s\\
    -\uv{y}\vdot\uv{e}_p&\uv{y}\vdot\uv{e}_s\\
    -\uv{z}\vdot\uv{e}_p&\uv{z}\vdot\uv{e}_s\\
    }\pmqty{A_s\\A_p}\,.
\end{equation}
This map can again be written in terms of Pauli matrices as:
\begin{equation}
    \pmqty{\uv{e}_s&\uv{e}_p\\-\uv{e}_p&\uv{e}_s}=\uv{e}_s\pmqty{\pmat{0}}+\ii\uv{e}_p \pmqty{0&-\ii\\\ii&0}=\uv{e}_s\tens{\sigma}_0+\ii\uv{e}_p \tens{\sigma}_2\,.
\end{equation}
We can define a left inverse of \(\tens{M}\) by requiring \(\tens{M}_L\tens{M}=\tens{\sigma}_0\) which is satisfied, for example, by:
\begin{equation}\label{eq:inverse}
    \tens{M}_L\tens{M}=\underbrace{\pmqty{\uv{e}^\intercal_s&0\\0&\uv{e}^\intercal_s}}_{2\times6}\underbrace{\pmqty{\uv{e}_s&\uv{e}_p\\-\uv{e}_p&\uv{e}_s}}_{6\times2}=\underbrace{\pmqty{1&0\\0&1}}_{2\times2}=\tens{\sigma}_0\,,\qq{since} \uv{e}^\intercal_s\uv{e}_s=1, \,\text{and }\uv{e}^\intercal_s\uv{e}_p=0\,,
\end{equation}
notice that there is not one unique inverse (a well-known feature of one-sided inverses), but infinitely many, as 
\begin{equation}
    [\tens{M}_L+\tens{A}(\tens{I}-\tens{M}\tens{M}_L)]\tens{M}=\tens{M}_L\tens{M}+\tens{A}(\tens{M}-\tens{M})=\tens{\sigma}_0\,,\quad \forall \tens{A}\in\mathbb{C}^{2\times6}\,,\;\tens{I}=\operatorname{diag}(1,1,1,1,1,1)\,.
\end{equation}
Using any of these inverses, we can project the electromagnetic bivector for any plane wave (or evanescent wave) onto the Jones vector as follows:
\begin{equation}
    \vb*{\mathcal{J}}=\tens{M}_L(\omega,\vc{k})c\vec{F}_0(\omega,\vc{k})\,,
\end{equation}
Now suppose that we want to know $\vb*{\mathcal{J}}'$ in a boosted frame, we can write
\begin{equation}
    \vb*{\mathcal{J}}'=\tens{M}'_L(\omega',\vc{k}')\operatorname{Ad}_{\Lambda}(\vcbeta)c\vec{F}_0(\omega,\vc{k})=\tens{M}'_L(\omega',\vc{k}')\operatorname{Ad}_{\Lambda}(\vcbeta)\tens{M}(\omega,\vc{k})\vb*{\mathcal{J}}=\tens{\Lambda}(\omega,\vc{k},\vcbeta)\vb*{\mathcal{J}}\,,
\end{equation}
where $\tens{\Lambda}(\omega,\vc{k},\vcbeta)$ is how the Lorentz boost is represented on the Jones vector {for any electromagnetic field with a $\ee^{\ii\vc{k}\vdot\vc{r}-\ii\omega t}$ dependence, including plane waves and evanescent waves}, accounting for the fact that the polarisation basis is different in each frame. We can use the \cref{eq:boost,eq:inverse} to write the most general expression for the boost of the Jones vector explicitly in terms of polarisation basis vectors as (choosing $\tens{M}'_L=\uv{e}'^\intercal_s\tens{\sigma}_0$):
\begin{equation}
    \tens{\Lambda}(\omega,\vc{k},\vcbeta)=\uv{e}'_s\vdot[\vc{\Gamma}_\perp\uv{e}_s-\gamma\vcbeta\cp\uv{e}_p]\tens{\sigma}_0+\uv{e}'_s\vdot[\vc{\Gamma}_\perp\uv{e}_p+\gamma\vcbeta\cp\uv{e}_s]\ii\tens{\sigma}_2\,.
\end{equation}
For a wavevector $\vc{k}^\pm=k_x\uv{x}+k_y\uv{y}\pm k_z\uv{z}$ with the $s$ and $p$ polarisation basis vectors defined in general as:
\begin{equation}
    \uv{e}_s=\frac{\uv{z}\cp\vc{k}_t}{k_t}=\frac{k_x\uv{y}-k_y\uv{x}}{k_t}\,,\quad
    \uv{e}^\pm_p=\frac{\pm\vc{k}_tk_z-k_t^2\uv{z}}{k_tk_0}=\frac{\pm(k_x\uv{x}+k_y\uv{y})k_z-k_t^2\uv{z}}{k_tk_0}\,,
\end{equation}
we can find the diagonal $\tens{\sigma}_0$ terms for an arbitrary boost $\vcbeta=\beta_x\uv{x}+\beta_y\uv{y}+\beta_z\uv{z}$ to be (after some algebra):
\begin{equation}
    \uv{e}'_s\vdot[\vc{\Gamma}_\perp\uv{e}_s-\gamma\vcbeta\cp\uv{e}^\pm_p]=\frac{\vc{k}_t'}{k_t'}\vdot\qty[\frac{\vc{k}_t'}{k_t}+\frac{\gamma k_z}{k_tk_0}\qty(\vcbeta_t{k_z}\mp\beta_z{\vc{k}_t})]=\frac{k'_t}{k_t}+\frac{k_z}{k_0}\frac{\gamma \qty(\vcbeta{k_z}\mp\beta_z{\vc{k}_t})\vdot\vc{k}_t'}{k_tk_t'}\,.
\end{equation}
Notice that $(\uvbeta\vdot\vc{k}_t')$ is the part of $\vc{k}_t'$ along the boost direction, 
so if $\uvbeta=\uv{x}$ we
obtain:
\begin{equation}
    \uv{e}'_s\vdot[\vc{\Gamma}_\perp\uv{e}_s-\gamma\beta\uv{x}\cp\uv{e}^\pm_p]=\frac{k'_t}{k_t}+\gamma\beta\frac{ k_x'k^2_z}{k_t'k_tk_0}\,,
\end{equation}
which is precisely the diagonal terms in \cref{eq:boost_of_Jones} in the main text.
The antisymmetric mixed terms (with $\ii\tens{\sigma}_2$) are
\begin{equation}
    \uv{e}'_s\vdot[\vc{\Gamma}_\perp\uv{e}^\pm_p+\gamma\vcbeta\cp\uv{e}_s]=
    \frac{(\uv{z}\cp\uvbeta)\vdot\vc{k}_t}{k'_tk_t}\qty[\frac{\mp\gamma{\beta}k'_0k_z}{k_0}+\frac{\beta_z({\gamma-1})}{\beta}\qty(\gamma\vcbeta\vdot\vc{k}'_t +(1+\gamma)k'_0-\frac{k_t^2}{k_0})]
\end{equation}
where $(\uv{z}\cp\uvbeta)\vdot\vc{k}_t=(\uv{z}\cp\uvbeta)\vdot\vc{k}'_t$ is the part of $\vc{k}_t$ that is both transverse to $\uvbeta$ as well as $\uv{z}$, hence for $\uvbeta=\uv{x}$:
\begin{equation}
    \uv{e}'_s\vdot[\vc{\Gamma}_\perp\uv{e}^\pm_p+\gamma\vcbeta\cp\uv{e}_s]=\mp\gamma\beta\frac{k'_0k_zk_y}{k_0k_t'k_t}\
\end{equation}
which matches the antisymmetric part in \cref{eq:boost_of_Jones} in the main text.

\bibliography{reference.bib}

\end{document}